\documentclass[twocolumn,english,superscriptaddress]{revtex4-1}
\usepackage[T1]{fontenc}
\usepackage[latin9]{inputenc}
\usepackage{graphicx}
\usepackage{amsmath}
\usepackage[colorlinks = true,
            linkcolor = blue,
            urlcolor  = blue,
            citecolor = blue,
            anchorcolor = blue]{hyperref}
\usepackage[usenames, dvipsnames]{color}

\makeatletter
\usepackage{babel}
\usepackage{float}

\makeatother

\begin{document}





\title{Crystal field Hamiltonian and anisotropy in $\rm KErSe_2$ and $\rm CsErSe_2$}

\author{A. Scheie}
\email{scheieao@ornl.gov}
\address{Neutron Scattering Division, Oak Ridge National Laboratory, Oak Ridge, Tennessee 37831, USA}

\author{V. O. Garlea}
\address{Neutron Scattering Division, Oak Ridge National Laboratory, Oak Ridge, Tennessee 37831, USA}

\author{L. D. Sanjeewa}
\address{Materials Science and Technology Division, Oak Ridge National Laboratory, Oak Ridge, Tennessee 37831, USA}

\author{J. Xing}
\address{Materials Science and Technology Division, Oak Ridge National Laboratory, Oak Ridge, Tennessee 37831, USA}

\author{A. S. Sefat}
\address{Materials Science and Technology Division, Oak Ridge National Laboratory, Oak Ridge, Tennessee 37831, USA}

\date{\today}

\begin{abstract}
We use neutron scattering and bulk property measurements to determine the single-ion crystal-field Hamiltonians of delafossites $\rm KErSe_2$ and $\rm CsErSe_2$. These two systems contains planar equilateral triangular Er lattices arranged in two stacking variants: rhombohedral (for K) or hexagonal (Cs). 
 Our analysis shows that regardless the stacking order both compound exhibit an easy-plane ground state doublet with large $J_z=1/2$ terms and the potential for significant quantum effects, making them candidates for quantum spin liquid or other exotic ground states. 
\end{abstract}

\maketitle

\section{Introduction}


The triangular lattice is a canonical geometry in theories of quantum spin liquids (QSL), wherein spins are entangled in a long-range fluctuating ground state with fractionalized excitations \cite{Knolle2019_review,Balents2010review,Savary_2016review}. Although isotropic quantum spins on a 2D triangular lattice order magnetically  \cite{Capriotti_1999,Huse_1988}, further neighbor exchange \cite{Zhu_2015_QSL,Zhu_2018} and magnetic anisotropy \cite{Zhou2017,Maksimov_2019} can theoretically produce a QSL state. 
Because magnetic anisotropy is driven by spin-orbit interactions, $f$-electron rare-earth ions are prime candidates for triangular-lattice QSLs \cite{Witczak-Krempa_2014,Li_2016,Iaconis_2018}. One of the most prominent rare-earth triangular lattice QSL candidates has been ${\mathrm{YbMgGaO}}_{4}$ \cite{Li_2015_YMGO,Li_2016_YMGO,Shen2016,Paddison2017,Xu_2016_YMGO}, although disorder-driven glassiness in the ground state casts doubt on the QSL hypothesis \cite{Li_2017_YMGO,Li_2018_YMGO,Zhu_2017_YMGO,Kimchi_2018,Zhen_2018}. Nevertheless, since the discovery of ${\mathrm{YbMgGaO}}_{4}$ there has been a salvo of rare-earth based triangular lattice QSL candidates \cite{Liu_2018_Chalcogenides,Baenitz_2018,Bordelon2019,Ranjinth2019,sarkar2019quantum,gao2019crystal,Ashtar2019,Ding_2019_NYO,Ranjith2019_2,Xing_2019_Field-induced}.
Any exotic behavior in these materials is heavily dependent upon magnetic anisotropy, so understanding the magnetic anisotropy is of key importance. 

Magnetic single-ion anisotropy  of an ion comes from crystal electric field (CEF) interactions with surrounding ligands \cite{Hutchings1964,EDVARDSSON1998}. The CEF Hamiltonian also determines the ease of quantum tunnelling of the effective spin ground state: a system with strong quantum tunnelling effects will have large  $| J_z=  \pm 1/2 \rangle$ coefficients in its ground state \cite{Rau_2015}. Transitions between CEF levels are visible in neutron scattering, which allows the CEF parameters to be fitted to the energies and intensities of these modes \cite{Furrer2009neutron}. 

\begin{figure}
	\centering\includegraphics[width=0.47\textwidth]{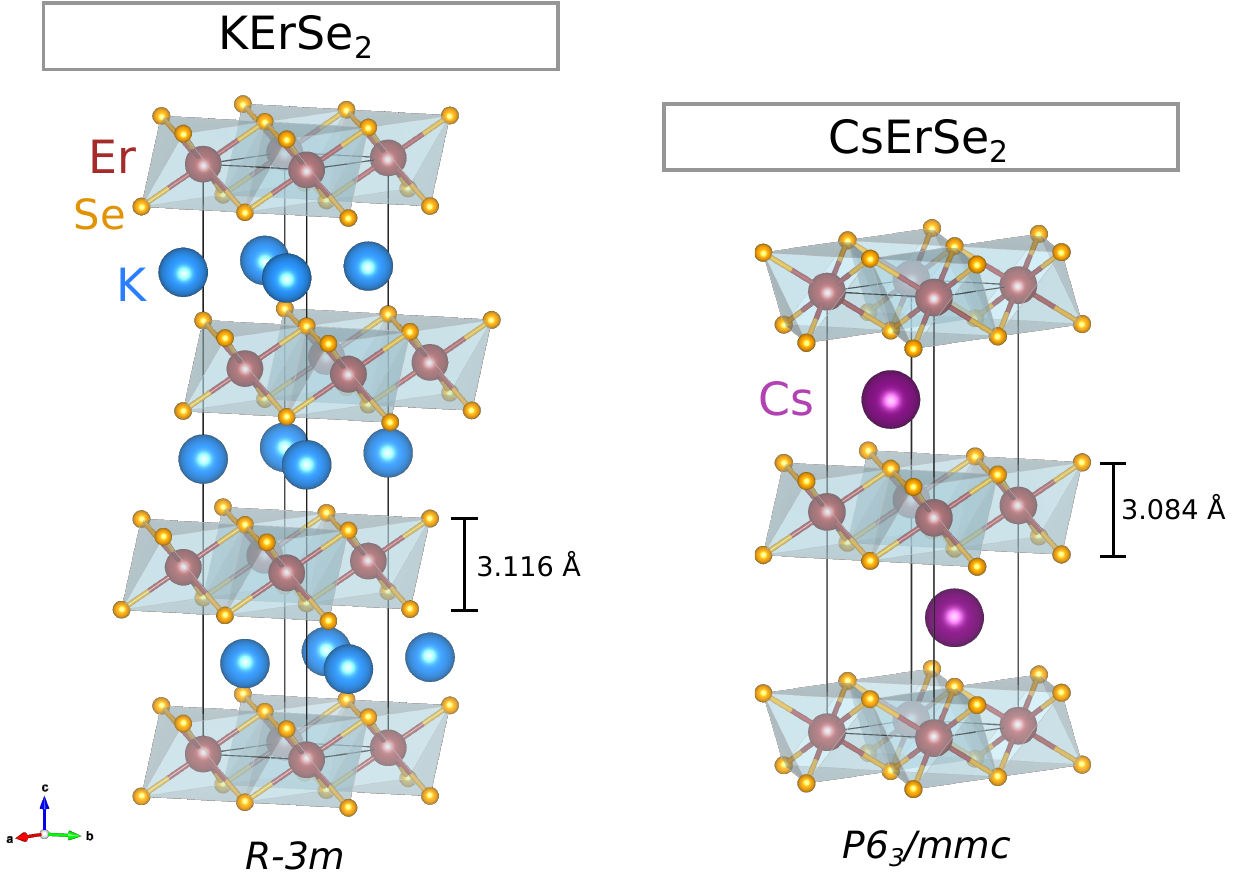}
	
	\caption{Crystal structures of  $\rm KErSe_2$ (left) and $\rm CsErSe_2$ (right), showing the different stacking of triangular Er lattices. The $\rm CsErSe_2$ Se octahedra are slightly compressed along the $c$ axis compared to $\rm KErSe_2$.}

	\label{flo:CrystStruct}
	
\end{figure}

Recently, a new family of rare-earth delafossite triangular lattice magnetic materials was reported based on the $AB{\rm Se_2}$ formula, where $A$ is an alkali ion and $B$ is a rare-earth ion \cite{xing2019_CES,xing2019_KES}. The whole series is triangular, but some of the compounds crystallize in the $R$-$3m$ space group while others crystallize in the $P6_3/mmc$ space group. This difference is in the stacking of triangular lattice layers, but it may also lead to subtle differences in magnetic anisotropy.
In this study, we use inelastic neutron scattering and magnetization to determine the magnetic anisotropy of Er$^{3+}$ triangular lattice materials $\rm KErSe_2$ ($R$-$3m$) \cite{xing2019_KES} and $\rm CsErSe_2$ ($P6_3/mmc$) \cite{xing2019_CES}, shown in Fig. \ref{flo:CrystStruct}. Both of these compounds show no magnetic order or spin freezing down to 0.42~K, and single-crystal magnetization shows an easy-plane magnetic anisotropy with low-field magnetization indicating a correlated magnetic state in $\rm KErSe_2$ \cite{xing2019_KES}.  For rare-earths, an easy-plane ansiotropy often indicates an effective $|J_z = \pm \frac{1}{2} \rangle$ ground state which allows for significant quantum effects. This, combined with the observed correlations and absence of magnetic order, makes these materials candidates for exotic magnetic behavior---possibly the spin liquid phase.
Our analysis confirms that the ground state doublet has a large $J=1/2$ contribution with the potential for appreciable quantum effects.

\section{Experiments}

Powder samples of  $\rm KErSe_2$ and $\rm CsErSe_2$  were synthesized via solid state synthesis under vacuum as described in refs. \cite{xing2019_KES}  and \cite{xing2019_CES} respectively; and single crystals were grown via KCl and CsCl flux as described in refs. \cite{xing2019_KES}  and \cite{xing2019_CES}.
We measured powder-average susceptibility at 1~T between 2~K and 300~K and single crystal magnetization at 2~K using a Quantum Design MPMS. For the single-crystal measurements, 0.56 mg ($\rm KErSe_2$) and 1.75~mg ($\rm CsErSe_2$) plate-like crystals were used, with field oriented along the $c$ axis (orthogonal to the plate surface) and then with field oriented in the $ab$ plane (parallel to the plate surface). 

We performed neutron scattering experiments using the HYSPEC instrument at the ORNL Spallation Neutron Source. Both $\rm KErSe_2$ and  $\rm CsErSe_2$ samples weighting approximately 3 grams were loaded in a loose powder form inside 9.5 mm diameter Aluminum cans. We measured the $\rm KErSe_2$ spectrum at 1.8~K, 15~K, 50~K and 100~K temperatures for 8 hours each with an incident neutron energy $E_i=9$~meV and Fermi chopper frequency of 360 Hz. Additional measurements were carried out at the same temperatures using $E_i=20$~meV. The sample was cooled down using a standard 100mm bore Orange cryostat. For the $\rm CsErSe_2$ compound, we collected data at $T=1.8$~K and $T=50$~K for 7.5 hours using $E_i=9$~meV and 360~Hz. Further measurements were carried out using $E_i=30$~meV at 1.8~K. This sample was mounted and cooled in a vertical field cryomagnet.  Measurements of  $\rm CsErSe_2$ under applied magnetic fields of up to 5 Tesla were performed to evaluate the Zeeman splitting of crystal field levels.  

\begin{figure}
	\centering\includegraphics[width=0.48\textwidth]{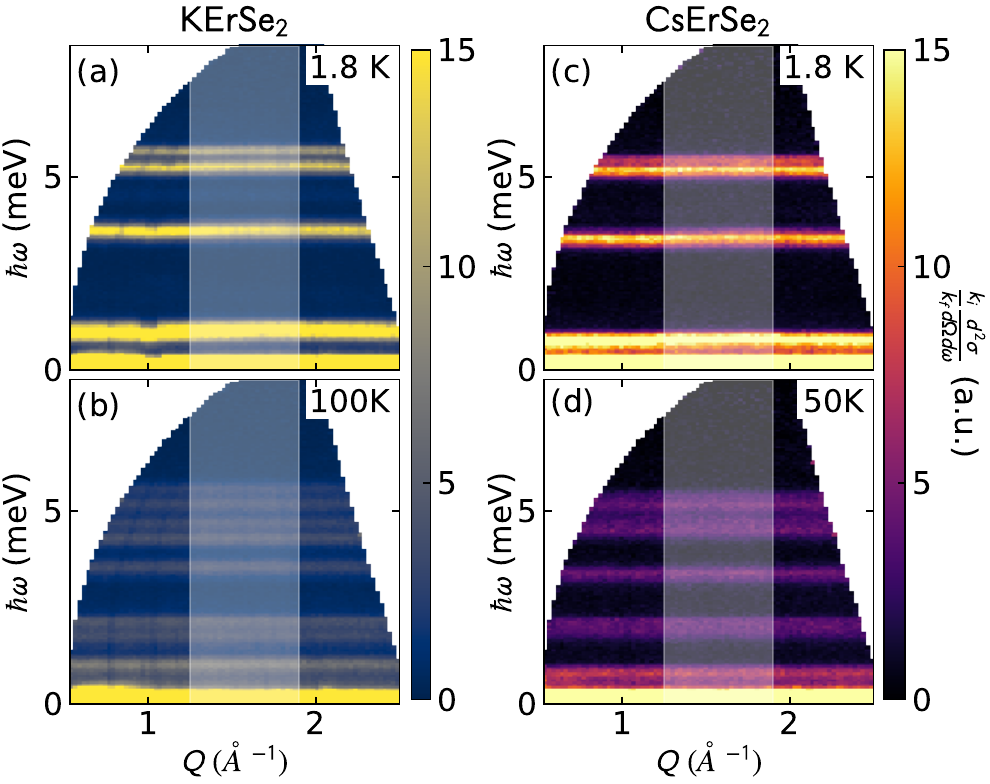}
	
	\caption{Powder neutron spectrum of $\rm KErSe_2$ at 1.8~K (a) and 100~K (b), and $\rm CsErSe_2$ at 1.8~K (c) and 50~K (d). The crystal field excitations are clearly visible. The grey region shows the $Q$ range used to fit the CEF Hamiltonian in Fig. \ref{flo:neutronfits}.}

	\label{flo:NeutronSpectrum}
	
\end{figure}

\section{Results}

The zero-field neutron scattering data for $\rm KErSe_2$ and $\rm CsErSe_2$ compounds measured using $E_i = 9$~meV is shown in Figs. \ref{flo:NeutronSpectrum} and \ref{flo:neutronfits}.
The neutron scattering spectra show clear CEF transitions in both compounds. At 1.8~K, $\rm KErSe_2$ has four visible modes at 0.915(6) meV, 3.504(9) meV, 5.15(1) meV, and 5.56(1) meV. Additional scans with $E_i=20$~meV revealed no additional peaks up to $\hbar \omega=19$~meV. Meanwhile, $\rm CsErSe_2$ has four visible modes at    0.731(8) meV, 3.34(1) meV, 5.10(2) meV,  and 5.38(3) meV.  Additional scans with $E_i=30$~meV revealed no additional peaks up to $\hbar \omega=29$~meV (see Appendix \ref{app:FittedData}).
At higher temperatures, the low-energy levels become populated and many more transitions are visible in the neutron spectrum. 

Low-temperature magnetization in Fig. \ref{flo:magnetization} shows that both $\rm KErSe_2$ and $\rm CsErSe_2$ are more easily magnetized along the in-plane direction than in the $c$ direction, indicating an easy-plane magnetic anisotropy.
Magnetic suscpetibility shows linear Curie-Weiss behavior for both materials, with $\mu_{eff} = 9.453(2) \> \mu_B$ for $\rm KErSe_2$ and $\mu_{eff} = 9.555(6) \> \mu_B$ $\rm CsErSe_2$ (fitted for 50~K$<T<$200~K). This is very close to the free-ion value of $9.581 \> \mu_B$. Fitted Weiss temperatures are not meaningful here because the low-lying CEF levels also induce an offset in the $x$ intercept which cannot be disentangled from mean-field exchange at the temperatures measured. Instead, we estimate the magnetic exchange ineraction with molecular field theory, see below.

The close correspondence between the experimental data for these compounds suggests that the CEF Hamiltonians of these two compounds are very similar. The challenge is fitting the data to the appropriate model.

\section{CEF Analysis}

The CEF Hamiltonian can be expressed using the Stevens Operator formalism as 
\begin{equation}
\mathcal{H}_{CEF} =\sum_{n,m} B_{n}^{m}O_{n}^{m}.
\end{equation}
Here $O_{n}^{m}$ are the Stevens Operators \cite{Stevens1952,Hutchings1964} and $B_{n}^{m}$ are multiplicative factors called CEF parameters. Er$^{3+}$ is a Kramers ion with an effective spin $J=15/2$, so up to eight doublet eigenstates will exist.
For both $\rm KErSe_2$ and $\rm CsErSe_2$, the Er$^{3+}$ ion has a $D_3$ symmetric ligand environment with a rotation axis about $c$. Setting the $z$ axis along $c$, symmetry dictates that only six CEF parameters are nonzero: $B_2^0$, $B_4^0$, $B_4^3$, $B_6^0$, $B_6^3$, and $B_6^6$ \cite{Hutchings1964}. These coefficients, once properly fitted to the data, uniquely define the CEF Hamiltonian.

To simplify the neutron data for the CEF fit, we integrated over $1.25$~\AA$^{-1} < Q < 1.9$~\AA$^{-1}$ to create 2D data sets.  This range was chosen to maximize the energy transfer range over which the same range of $Q$ is integrated. Given that the CEF excitations are local and have no dispersion, no information is lost by doing this.  For $\rm KErSe_2$, we simultaneously fit the CEF model to the 1.8 K, 15 K, and 100 K data. For $\rm CsErSe_2$, we simultaneously fit the CEF model to the 1.8~K and 50~K data.


We fit the CEF Hamiltonian directly to the measured neutron spectra, rather than extracting peak energies and intensities beforehand. Thus we avoid making assumptions about overlapping peaks in the high-temperature data sets.
For both  $\rm KErSe_2$ and $\rm CsErSe_2$, we defined the starting parameters with a point-charge model wherein surrounding ligands are modeled as electrostatic point-charges \cite{Hutchings1964}.  We then fit the effective positions of the ligands to the neutron data, and then used the CEF parameters from that intermediate fit as starting values for fitting the CEF parameters directly. All fits and CEF calculations were performed using PyCrystalField software \cite{PyCrystalField}, and details of the fitting procedure are in Appendix \ref{app:FitProtocol}.

\begin{figure*}
	\centering\includegraphics[width=0.99\textwidth]{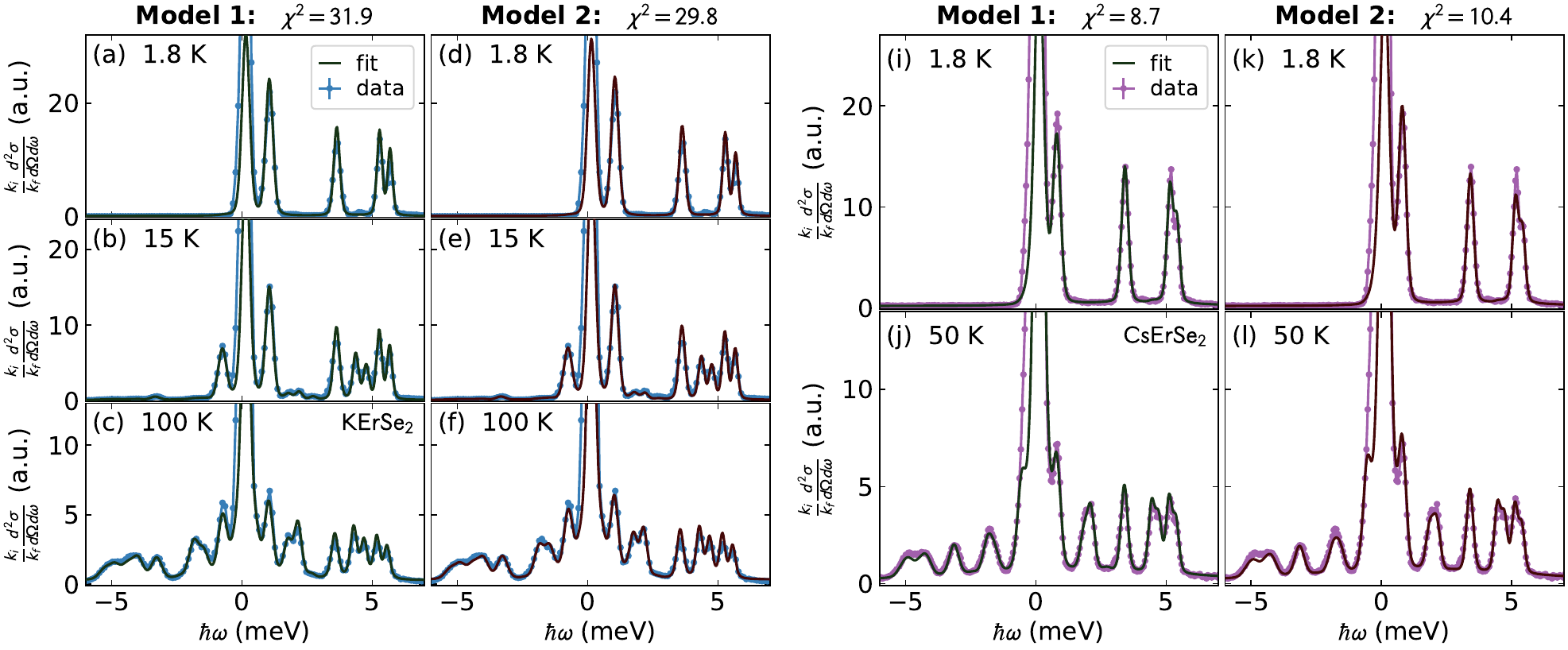}
	
	\caption{Crystal field fits to neutron scattering data for $\rm KErSe_2$  (a) - (f) and $\rm CsErSe_2$ and  (i) - (l). Each column shows the fit for a different model, both of which fit the data well.}

	\label{flo:neutronfits}
	
\end{figure*}

The initial fitted CEF Hamiltonians yielded excellent matches to the neutron scattering data, but we found two different models which fit the $\rm KErSe_2$ and $\rm CsErSe_2$ neutron scattering data. One model shows an  easy-plane ground state with $B_2^0 < 0$, the other shows an easy-axis ground state with $B_2^0 > 0$, with small variations between each material in the CEF parameters.   
The easy-plane model we call \textit{Model 1} and the easy-axis model we call \textit{Model 2}. Both Model 1 and Model 2 fits are shown in Fig. \ref{flo:neutronfits}.
Assuming that similar chemical structures will lead to similar CEF Hamiltonians, only one model is correct. To adjudicate, we turn to bulk property measurements.

\begin{figure*}
	\centering\includegraphics[width=0.99\textwidth]{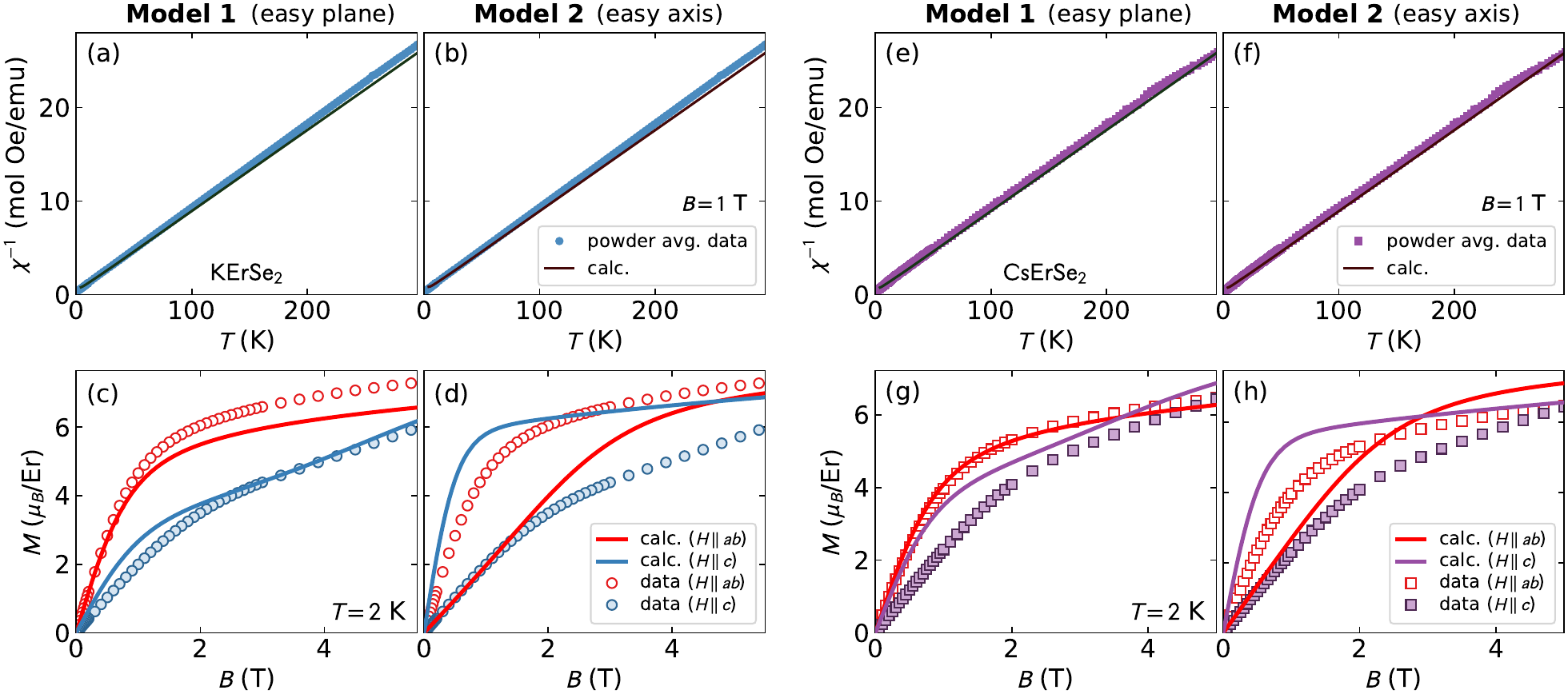}
	
	\caption{Calculated powder-average susceptibility and oriented magnetization for $\rm KErSe_2$  (a) - (d) and $\rm CsErSe_2$ and  (e) - (h). For both compounds, the calculated powder-average susceptibility from Model 1 and Model 2 are indistinguishable. However, the low-temperature magnetization is very different: Model 1 predicts an easy-plane anisotropy, while Model 2 predicts an easy-axis. In both cases, Model 1 provides a better match for the data.}

	\label{flo:magnetization}
	
\end{figure*}

We computed the powder-averaged single-ion susceptibility from the CEF Hamiltonians for $\rm KErSe_2$ [Fig. \ref{flo:magnetization}(a)-(b)] and $\rm CsErSe_2$ [Fig. \ref{flo:magnetization}(e)-(f)], but the powder-averaged susceptibilities for Models 1 and 2 are nearly identical for both compounds: the $\chi^2$ differ by less than 0.2\%.   Thus it is not possible to distinguish between the two models with powder susceptibility.

Fortunately, Models 1 and 2 can be distinguished with directional magnetization, which clearly shows both  $\rm KErSe_2$ and $\rm CsErSe_2$ to have an easy-plane orthogonal to the $c$ direction.
We computed the directional magnetization from the CEF Hamiltonians for $\rm KErSe_2$ [Fig. \ref{flo:magnetization}(c)-(d)] and $\rm CsErSe_2$ [Fig. \ref{flo:magnetization}(g)-(h)]. Model 1 magnetization reveals an easy-plane ground state, while Model 2 magnetization reveals an easy-axis. Therefore, we  identify Model 1 as the correct CEF model for both $\rm KErSe_2$ and $\rm CsErSe_2$. 
The correspondence between experimental and theoretical magnetization curves is not perfect because magnetic exchange interactions severely affect the shape of magnetization curves at low temperatures. Nevertheless, the overall anisotropy is clear.

\begin{table}
\caption{Best fit (Model 1) CEF parameters for $\rm{KErSe_2}$ and $\rm{CsErSe_2}$. Four significant figures are given for each value regardless of uncertainty so that the CEF levels can be reproduced.}
\begin{ruledtabular}
\begin{tabular}{c|cc}
$B_n^m$ (meV) &$\rm{KErSe_2}$  &  $\rm{CsErSe_2}$ \tabularnewline
 \hline 
$ B_2^0$ & $ (-2.773 \pm 0.33)\times 10^{-2}$ & $ (-3.559 \pm 0.64)\times 10^{-2}$ \tabularnewline
$ B_4^0$ & $ (-3.987 \pm 0.05)\times 10^{-4}$ & $ (-3.849 \pm 0.11)\times 10^{-4}$ \tabularnewline
$ B_4^3$ & $ (-1.416 \pm 0.02)\times 10^{-2}$ & $ (-1.393 \pm 0.03)\times 10^{-2}$ \tabularnewline
$ B_6^0$ & $ (3.152 \pm 0.02)\times 10^{-6}$ & $ (3.154 \pm 0.04)\times 10^{-6}$ \tabularnewline
$ B_6^3$ & $ (-7.616 \pm 1.94)\times 10^{-6}$ & $ (-4.695 \pm 3.56)\times 10^{-6}$ \tabularnewline
$ B_6^6$ & $ (3.275 \pm 0.19)\times 10^{-5}$ & $ (3.381 \pm 0.37)\times 10^{-5}$ \tabularnewline
\end{tabular}\end{ruledtabular}
\label{tab:CES+KES_CEF_params}
\end{table}

The best fit CEF parameters, taken from Model 1, are given in Table \ref{tab:CES+KES_CEF_params}. 
\begin{widetext}
The lowest energy doublet for $\rm KErSe_2$ is
\begin{equation}
     |\psi_{\pm} \rangle = \pm 0.52(2) \Big| \mp \frac{13}{2}\Big\rangle -0.508(5) \Big| \mp \frac{7}{2}\Big\rangle \pm 0.58(3) \Big| \mp \frac{1}{2}\Big\rangle+0.347(6) \Big| \pm \frac{5}{2}\Big\rangle \pm 0.118(6) \Big| \pm \frac{11}{2}\Big\rangle \label{eq:GS1} ,
\end{equation}
and the lowest energy doublet for  $\rm CsErSe_2$ is
\begin{equation}
    |\psi_{\pm} \rangle = \pm 0.59(4) \Big| \mp \frac{13}{2}\Big\rangle   +0.513(3) \Big| \mp \frac{7}{2}\Big\rangle 
    \pm 0.51(5) \Big| \mp \frac{1}{2}\Big\rangle -0.338(2) \Big| \pm \frac{5}{2}\Big\rangle \pm 0.123(9) \Big| \pm \frac{11}{2}\Big\rangle  \label{eq:GS4}.
\end{equation}
\end{widetext}
The full lists of eigenvalues and eigenstates are given in Tables \ref{tab:KES_Eigenvectors} and \ref{tab:CES_Eigenvectors}. The $g$ tensors calculated from the ground state kets are $g_{\perp}=6.0(1)$ and $g_{z}=4.9(2)$ for  $\rm KErSe_2$; and $g_{\perp}=5.4(3)$ and $g_{z}=5.9(5)$ for  $\rm CsErSe_2$. For  $\rm KErSe_2$ the $g$ tensor is easy-plane, but for $\rm CsErSe_2$ both easy-axis and easy-plane $g$-tensors are within uncertainty. Both are in qualitative agreement with the anisotropy indicated by low-field magnetization. 

\begin{figure}
	\centering\includegraphics[width=0.47\textwidth]{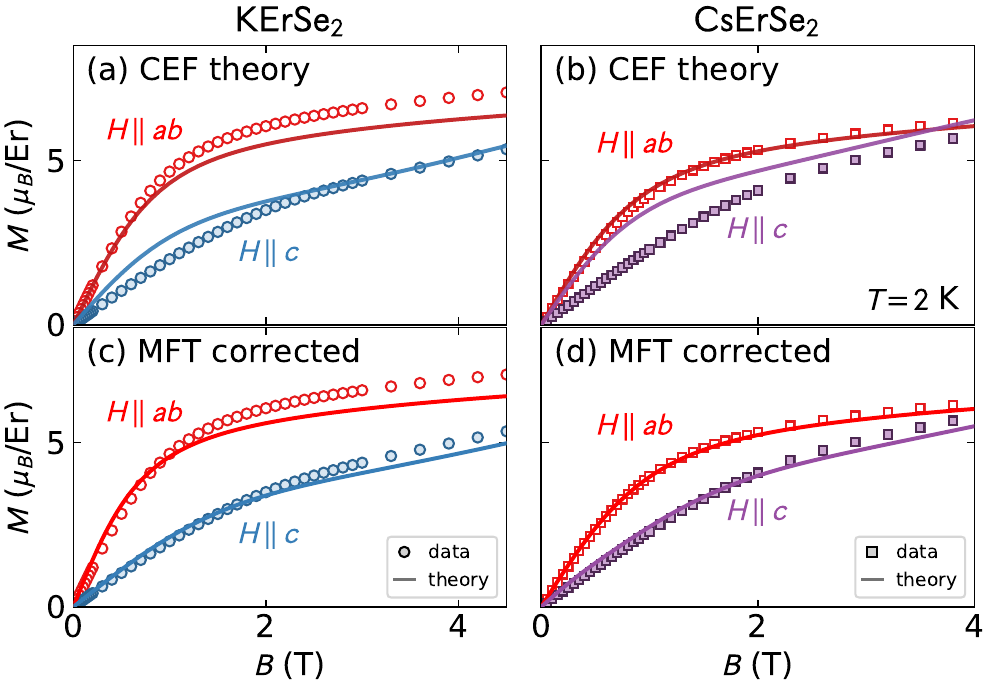}
	
	\caption{$\rm KErSe_2$ and $\rm CsErSe_2$ magnetization compared to the single-ion CEF calculations (a)-(b) and molecular field theory (MFT) corrected CEF calculated magnetization (c)-(d), which is used to estimate the strength of magnetic exchange. For both compounds, the $ab$-plane correction is small, but the $c$-axis correction is substantial.}
	
	\label{flo:MFT-magnetization}
	
\end{figure}

We improve agreement with experimental magnetization by incorporating exchange effects with molecular field theory. Assuming an effective field $H_{eff} = H_{ext}+\lambda M(H)$ where $H_{ext}$ is the external magnetic field and $\lambda = \frac{N {\mathcal J}}{(\mu_B g_J)^2}$ ($N$ is the number of nearest neighbors, $\mathcal J$ is the molecular field exchange), we can estimate the strength of the in-plane and out-of-plane magnetic exchange by fitting ${\mathcal J}$ to a molecular-field corrected CEF magnetization calculation. The fits are shown in Fig. \ref{flo:MFT-magnetization}.
 For $\rm KErSe_{2}$, fitted ${\mathcal J}_{ab} = 0.4(3) \> {\rm \mu eV}$ and ${\mathcal J}_{c} = -1.8(1.0) \> {\rm \mu eV}$. For $\rm CsErSe_{2}$, fitted ${\mathcal J}_{ab} = -0.2(6) \> {\rm \mu eV}$ and ${\mathcal J}_{c} = -2.4(5) \> {\rm \mu eV}$.
 These exchange constants are tiny (particularly the in-plane exchange). This is partly due to the very large effective spin: treating the Er$^{3+}$ $J=15/2$ as effective $S=1/2$  would give exchange 
$\frac{J(J+1)}{S(S+1)} = 85$ times greater ($\sim 0.2$~meV).
The large difference between in-plane and out-of-plane exchange indicates highly anisotropic magnetic exchange interactions as is common for rare earth ions. 

The uncertainties for all values were calculated by finding a line through parameter space which minimizes reduced $\chi^2$ up to one standard deviation from the global minimum (see Appendix \ref{app:uncertainty} for details).

\section{Discussion}

Both $\rm{KErSe_2}$  and  $\rm{CsErSe_2}$ have an easy-plane magnetic anisotropy coinciding with the triangular lattice plane, just like their Yb$^{3+}$ cousins \cite{Zangeneh2019,Sichelschmidt_2019}.
For both compounds, the ground state doublet has substantial weight given to $| \pm \frac{1}{2} \rangle$, $| \pm \frac{7}{2} \rangle$, and $| \pm \frac{13}{2} \rangle$. 
The similarity between the CEF ground states of these compounds shows that the difference between $R$-$3m$ and $P6_3/mmc$ does not produce a significant difference in anisotropy. Despite the different inter-layer arrangement of K and Cs ions, and the fact that the Se-octahedra of $\rm{KErSe_2}$  are 0.032(5) \AA~ taller along the $c$ axis with an Er-Se distance 0.008(7) \AA~ shorter, the components of the ground state eigenstates almost overlap to within uncertainty.

The large $| \pm \frac{1}{2} \rangle$ component in the CEF ground state means that $J_+$ and $J_-$ will have a significant effect in causing tunnelling between these two states, but the $| \pm \frac{7}{2} \rangle$ and $| \pm \frac{13}{2} \rangle$ could suggest more classical behavior. 
Thus, an exotic finite-temperature quantum state like a QSL is a real possibility but not guaranteed.


Easy-plane Er$^{3+}$ magnetism seems to be a robust feature of the delafossites: 
the $\rm{KErSe_2}$  and  $\rm{CsErSe_2}$ CEF Hamiltonians are very similar to the CEF ground state reported for triangular-lattice $\rm NaErS_2$, which also has easy-plane Er$^{3+}$ ground state with substantial weight on $| \pm \frac{1}{2} \rangle$ \cite{gao2019crystal}.
This is also similar to Er$^{3+}$ pyrochlores ${\rm Er}_2B_2{\rm O}_7$ ($B=$ Ge, Ti, Pt, and Sn) which likewise have easy-plane magnetic anisotropies  from a $D_3$ CEF environment \cite{Gaudet_2018}.
In the pyrochlore $\rm Er_2Ti_2O_7$, this easy-plane anisotropy leads to a degenerate ground state with emergent "clock anisotropies" in its magnetic ground state \cite{Ross_2014, Gaudet_2017,Zhitomirsky_2014}. Given the similar $XY$ CEF Hamiltonians, similar such behaviors could be found in the 2D triangular lattice Er$^{3+}$ delafossites. Quantum order by disorder, which is theorized to govern $\rm Er_2Ti_2O_7$ \cite{Ross_2014}, is also expected for triangular lattices \cite{Lecheminant_1995}, raising the possibility of emergent degeneracies on a 2D triangular lattice.

It is also worth noting that excited CEF states of $\rm{KErSe_2}$  and  $\rm{CsErSe_2}$  are at very low energy, so we expect them to influence magnetic exchange interactions via virtual crystal field fluctuations \cite{Petit_2014,Rau_2016}. Accordingly, the crystal field eigenstates presented here will be relevant to future theoretical investigations of these compounds.
To obtain more information on magnetic exchange, it will be necessary to measure at lower temperatures to ascertain whether these materials order magnetically and if so what type. 

\section{Conclusion}

We have used crystal field excitations and bulk magnetization to determine the crystal field ground state and anisotropy of $\rm{KErSe_2}$  and  $\rm{CsErSe_2}$, both of which have easy-plane ground state doublets despite the different crystal space groups. We report the full CEF Hamiltonian, which has significant $J_z=\frac{1}{2}$ components in the ground state doublet.


These results suggest that quantum effects are significant in the Er-based delafossites at low temperatures, making them candidates for quantum effects or emergent degeneracies like $\rm Er_2Ti_2O_7$. Additional low-temperature data is necessary to determine what, if any, is the ground state magnetic order. 
The lack of magnetic order and potential for strong quantum effects makes  $\rm{KErSe_2}$  and  $\rm{CsErSe_2}$  candidates for exotic magnetic states.

\subsection*{Acknowledgments}
The research is supported by the U.S. Department of Energy (DOE), Office of Science, Basic Energy Sciences (BES), Materials Science and Engineering Division.
This research used resources at the Spallation Neutron Source, a DOE Office of Science User Facility operated by the Oak Ridge National Laboratory. We also acknowledge helpful suggestions with Steve Nagler on fitting and statistics.

\quad


\appendix

\section{Fitted Data}\label{app:FittedData}

\begin{figure}
	\centering\includegraphics[width=0.47\textwidth]{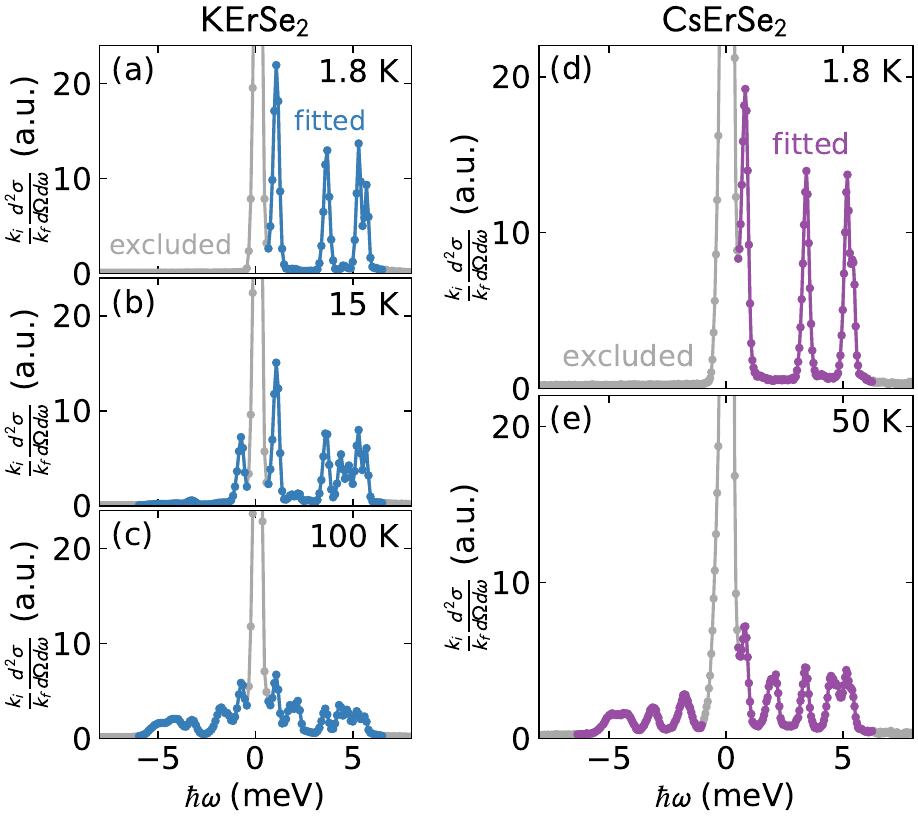}
	
	\caption{Data used to fit  $\rm KErSe_2$ (a)-(c) and $\rm CsErSe_2$ (d)-(e). Fitted data points are in color, excluded data points are in grey.}

	\label{flo:FittedData}
	
\end{figure}

We created 2D data sets to simplify the fitting procedure by integrating over $1.25$~\AA$^{-1} < Q < 1.9$~\AA$^{-1}$. We further simplified the data by excluding certain regions from the fitted data as shown in Fig. \ref{flo:FittedData}. The central elastic peak was excluded, as was the highest and lowest energy transfer data, which are featureless. We also excluded the negative energy transfer data for 1.8 K data because the Boltzmann population factor suppresses the negative energy peaks and there is no information there. At higher temperatures the negative energy peaks are visible, so we kept these in the range of fitted data.

\begin{figure}
	\centering\includegraphics[width=0.45\textwidth]{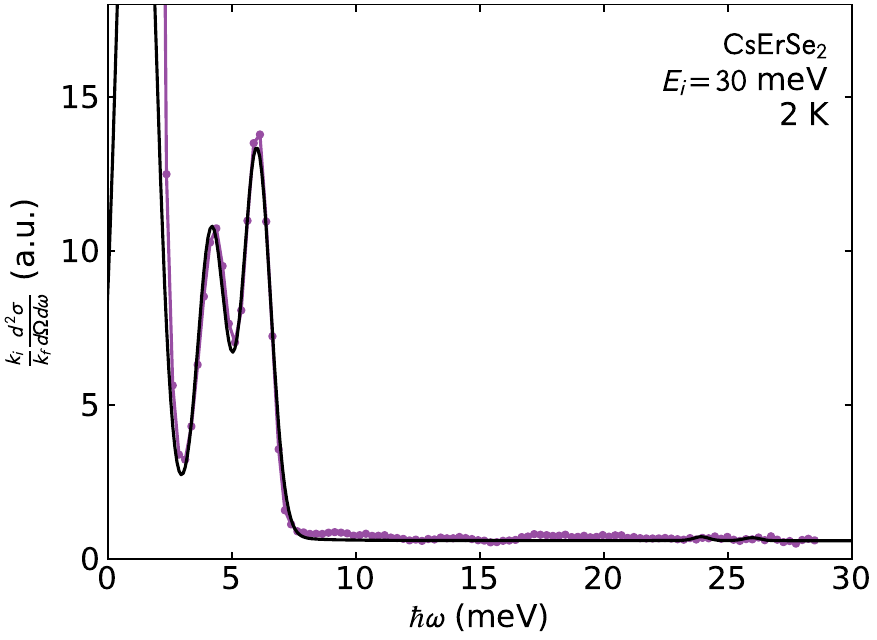}
	
	\caption{Observed and calculated scattering for $\rm CsErSe_2$ with $E_i = 30$ meV. Three doublets exist at 23.0 meV, 24.9 meV, and 25.2 meV in Model 1, but the calculated scattering is too weak to be observed. This is consistent with the data.}

	\label{flo:CES_30meV}
	
\end{figure}

We also collected neutron scattering data at $E_i = 30$ meV for $\rm CsErSe_2$, shown in Fig. \ref{flo:CES_30meV}. No peaks were visible in the data, and because of this the data was not used in the fits. Both Model 1 and Model 2 predict three CEF transitions around 25 meV, but for neither model do the peaks have any appreciable intensity. Model 1 calculated intensity is shown in Fig. \ref{flo:CES_30meV}, and no calculated peaks are visible, consistent with the data.

\section{Fitting procedure}\label{app:FitProtocol}


\subsection{Fitted variables}

In addition to the six CEF parameters, we included several fitted parameters in order to properly model the neutron spectrum. We fitted an overall intensity factor (different for each compound). We also included a linear fitted background in $\rm{KErSe_2}$. The background in $\rm{CsErSe_2}$ was larger and more complex because of the magnet used in the experiment, so we modeled the background with two broad Gaussians, adjusted by hand so that the background in between the CEF peaks matched experiment.

To model the peak shape, we used a Voigt profile to simulate a convoluted Gaussian and Lorentzian. The Gaussian component was defined by a phenomenological resolution function which models resolution width as a linear function of energy, defined by the widths of the 1.8~K peak widths for each compound. The Lorentzian component was treated as a fitted variable, constant as a function of energy but variable with temperature. We also fitted a global offset in energy for each compound to account for slight asymmetries in the resolution function. 

To account for thermal expansion shifting the ligand octahedra, we added a fitted thermal expansion parameter $E$ which multiplies the CEF transition energy by a factor varying linearly with temperature $\Delta = \Delta_{calc}(1 - E T)$ where $\Delta$ is the peak energy and $T$ is temperature.

In total, this gives 15 fitted parameters for $\rm{KErSe_2}$ and 11 fitted parameters for $\rm{CsErSe_2}$. The peak width, background, energy offset, and thermal expansion parameters were only fitted in the final stages---once the peak energies and intensities had converged roughly to their experimental values.

\subsection{Fitting protocol}

Following the method in ref. \cite{Scheie_2018}, we first fitted a point charge model and then directly fitted the CEF parameters.
We fitted the point charge model by varying the size of the Se octahedra and the compression along the $c$ axis (the only ways to modify the ligand environment while preserving all symmetries). Then we fit the CEF parameters directly to the data iteratively using the Powell \cite{PowellsMethod} and Nelder-Mead \cite{Nelder-Mead} methods. 
For initial fit stages, we added a term to the global $\chi^2$ based off the lowest-temperature transitions $\chi^2_{energy} = \sum_i (\Delta_{calc_i} - \Delta_{exp_i})^2$
to ensure the peaks converged to the right energies. 

The code for these fits, which gives the precise protocols and all intermediate fitted values, can be found at \href{https://github.com/asche1/PyCrystalField/tree/master/AErSe2}{https://github.com/asche1/PyCrystalField}.

\subsection{Fit results}\label{app:fitresults}

The fitted CEF parameters for Model 1 and Model 2 for $\rm{KErSe_2}$ are listed in Table \ref{tab:KES_CEF_params} and  for $\rm{CsErSe_2}$ are listed in Table \ref{tab:CES_CEF_params}. 
The reduced $\chi^2$ for $\rm KErSe_2$ is slightly lower for Model 2 ($\chi^2_{red_{M1}} = 31.9$ vs $\chi^2_{red_{M2}} = 29.8$), while reduced $\chi^2$ for $\rm CsErSe_2$ is slightly lower for Model 1 ($\chi^2_{red_{M1}} = 8.7$ vs $\chi^2_{red_{M2}} = 10.4$).  The overall $\chi^2_{red}$ is larger for $\rm KErSe_2$ than $\rm CsErSe_2$, possibly because of the more sophisticated background modeling used for $\rm CsErSe_2$.
Meanwhile, for $\rm KErSe_2$ inverse susceptibility 10~K$<T<300$~K, $\chi^2_{M1} = 349.6$ and $\chi^2_{M2} = 349.4$; for $\rm CsErSe_2$, $\chi^2_{M1} = 54.5$ and $\chi^2_{M2} = 54.4$.

\begin{table}
	\caption{Fitted CEF parameters for $\rm{KErSe_2}$}
	\begin{ruledtabular}
		\begin{tabular}{c|cc}
			$B_n^m$ (meV) &Model 1 & Model 2 \tabularnewline
			\hline 
			$ B_2^0$ & -2.773$\times 10^{-2}$ & 2.720$\times 10^{-2}$ \tabularnewline
			$ B_4^0$ & -3.987$\times 10^{-4}$ & -4.864$\times 10^{-4}$ \tabularnewline
			$ B_4^3$ & -1.416$\times 10^{-2}$ & 1.282$\times 10^{-2}$ \tabularnewline
			$ B_6^0$ & 3.152$\times 10^{-6}$ & 1.028$\times 10^{-6}$ \tabularnewline
			$ B_6^3$ & -7.616$\times 10^{-6}$ & 4.764$\times 10^{-5}$ \tabularnewline
			$ B_6^6$ & 3.275$\times 10^{-5}$ & 2.113$\times 10^{-5}$ \tabularnewline
	\end{tabular}\end{ruledtabular}
	\label{tab:KES_CEF_params}
\end{table}

\begin{table}
	\caption{Fitted CEF parameters for $\rm{CsErSe_2}$}
	\begin{ruledtabular}
		\begin{tabular}{c|cc}
			$B_n^m$ (meV) &Model 1 & Model 2 \tabularnewline
			\hline 
			$ B_2^0$ & -3.559$\times 10^{-2}$ & 3.114$\times 10^{-2}$ \tabularnewline
			$ B_4^0$ & -3.849$\times 10^{-4}$ & -4.718$\times 10^{-4}$ \tabularnewline
			$ B_4^3$ & -1.393$\times 10^{-2}$ & 1.259$\times 10^{-2}$ \tabularnewline
			$ B_6^0$ & 3.154$\times 10^{-6}$ & 9.324$\times 10^{-7}$ \tabularnewline
			$ B_6^3$ & -4.695$\times 10^{-6}$ & 4.715$\times 10^{-5}$ \tabularnewline
			$ B_6^6$ & 3.381$\times 10^{-5}$ & 2.011$\times 10^{-5}$ \tabularnewline
	\end{tabular}\end{ruledtabular}
	\label{tab:CES_CEF_params}
\end{table}

\section{Finite-field scattering}\label{app:finitefield}

\begin{figure}[H]
	\centering\includegraphics[width=0.44\textwidth]{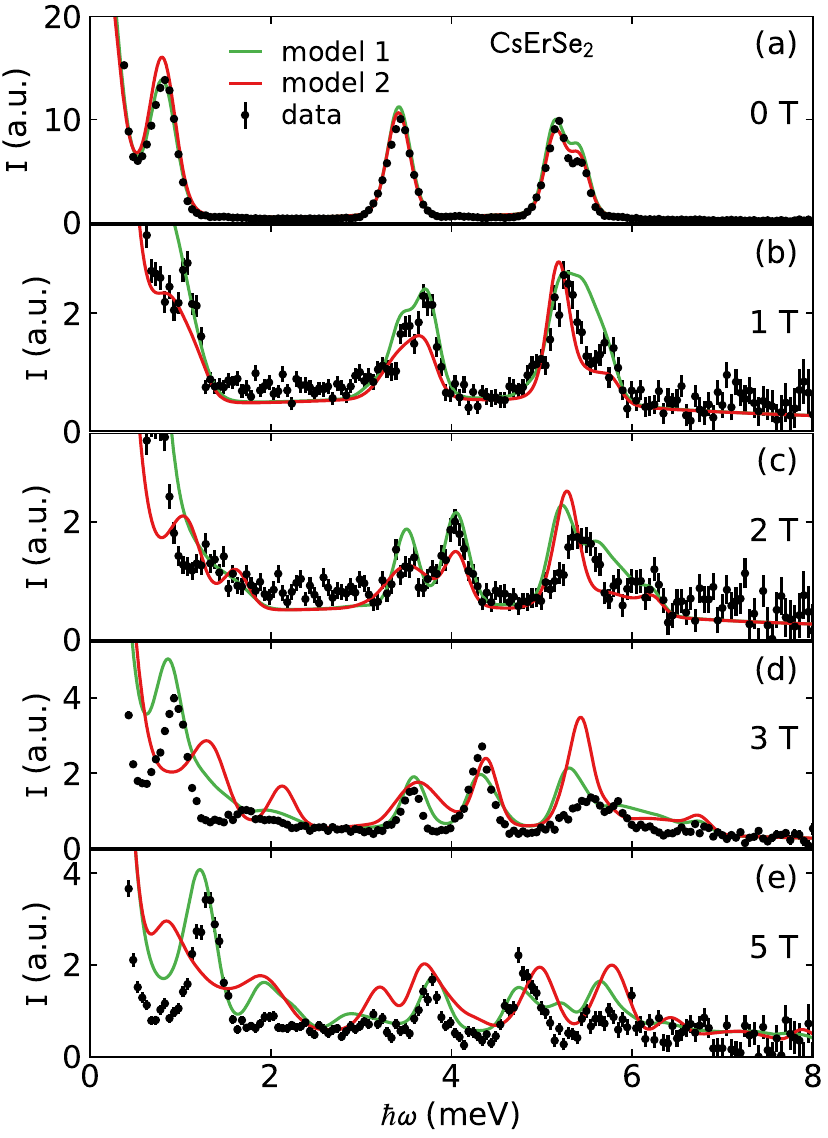}
	
	\caption{Nonzero-field CEF scattering for $\rm CsErSe_2$. Simulated scattering from Model 1 and Model 2 assuming a powder average are plotted in green and red, respectively. Neither model is perfect due to the imperfect powder average, but Model 1 is qualitatively closer to the measured scattering.}
	
	\label{flo:CES_fieldScan}
	
\end{figure}

We performed the $\rm CsErSe_2$ experiment with a vertical field magnet, and we collected data at 3~T, 5~T, 1~T and 2~T (in that order) in addition to 0~T. This data is shown in Fig. \ref{flo:CES_fieldScan}, and shows the doublets being split by the magnetic field.
Applying the magnetic field dramatically attenuated the neutron signal from the material (but not the background), indicating that the loose powder grains were shifted partially out of the beam by the magnetic field. This shifting almost certainly involved a reorientation in the powder grains---meaning that the magnetic field was preferentially applied along certain crystallographic directions.

The shifts and reorientation notwithstanding, we simulated the powder-average in-field neutron spectrum using PyCrystalField with field directions randomly sampled around a unit sphere. These simulations are plotted in Fig. \ref{flo:CES_fieldScan} for Model 1 (green) and Model 2 (red). 
The match between theory and experiment is not perfect, indicating the effects of grain reorientation, small magnetoelastic effects, or an slightly inaccurate Hamiltonian. Nevertheless, on a qualitative level, Model 1 matches the data much better than Model 2---particularly in the splitting of the low-energy mode at 3~T and 5~T.

\section{Uncertainty}\label{app:uncertainty}

\begin{figure*}
	\centering\includegraphics[width=0.99\textwidth]{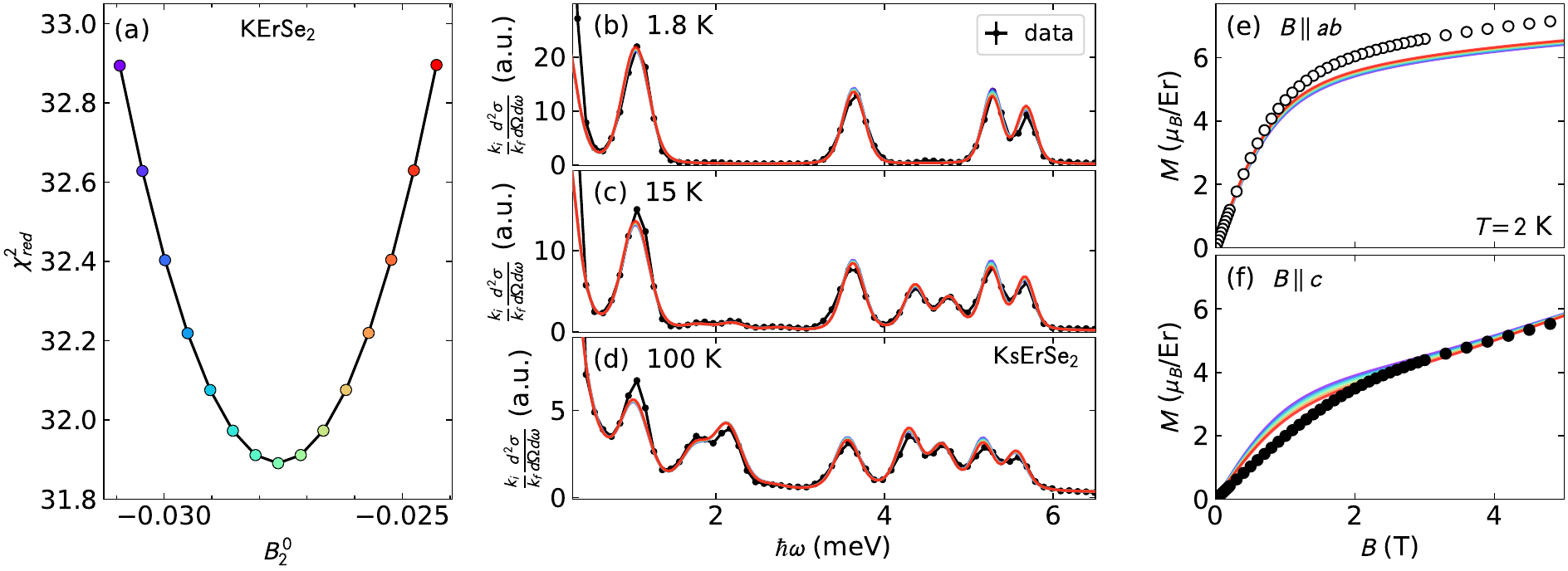}
	
	\caption{Uncertainty in fitted $\rm KErSe_2$ CEF parameters for Model 1. (a) $\chi^2_{red}$ as a function of $B_2^0$, allowing all other parameters to vary, up to one standard deviation.  (b)-(d) Calculated neutron spectrum for the range of $B_2^0$, with $B_2^0$ value indicated by the colors in panel (a). The various curves are nearly indistinguishable. (e)-(f) Magnetization for the  range of $B_2^0$.}

	\label{flo:KES_rangeOfSolutions}
	
\end{figure*}

\begin{figure*}
	\centering\includegraphics[width=0.99\textwidth]{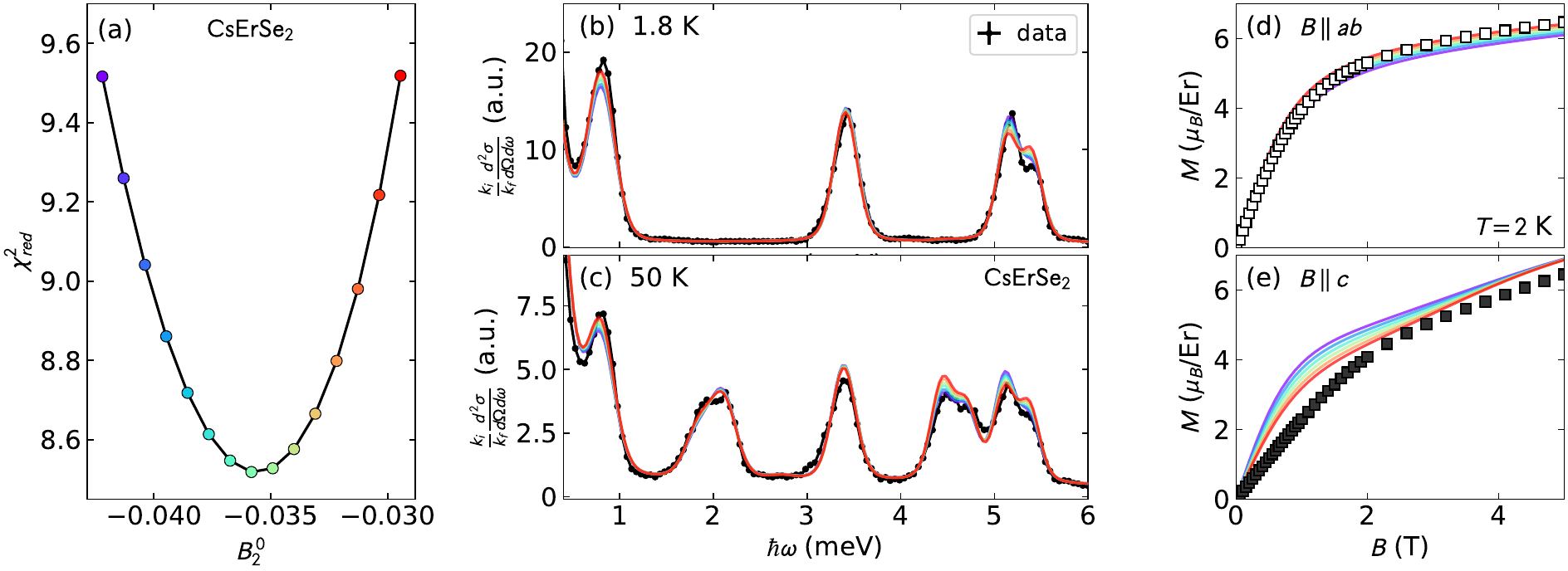}
	
	\caption{Uncertainty in fitted $\rm CsErSe_2$ CEF parameters for Model 1. (a) $\chi^2_{red}$ as a function of $B_2^0$, allowing all other parameters to vary, up to one standard deviation.  (b)-(c) Calculated neutron spectrum for the range of $B_2^0$, with $B_2^0$ value indicated by the colors in panel (a). The various curves are nearly indistinguishable. (d)-(e) Magnetization for the  range of $B_2^0$.}

	\label{flo:CES_rangeOfSolutions}
	
\end{figure*}

To characterize the uncertainty of the fitted CEF Hamiltonian, we 
defined a range of $B_2^0$ values around the best fit $B_2^0$ value, and for each $B_2^0$ we re-fit the neutron data varying all other variables. This resulted in a range of solutions which fit the data approximately equally.  These solutions, with the associated reduced $\chi^2$ and comparison to magnetization, are plotted in Fig. \ref{flo:KES_rangeOfSolutions} ($\rm KErSe_2$) and Fig. \ref{flo:CES_rangeOfSolutions} ($\rm CsErSe_2$).
(Differences in $c$-axis single-ion low-field magnetization are visible, but due to the strong non-trivial effects of magnetic exchange on low-temperature magnetization, we did not include magnetization in the global $\chi^2$ calculations.)

The range of solutions where the increase in $\chi^2$ is less than one above the minimum value gives us the uncertainty in the CEF parameters $B_n^m$ reported in Table \ref{tab:CES+KES_CEF_params} and the ground state kets in Eqs. \ref{eq:GS1}-\ref{eq:GS4}. 
The full list of eigenstates with associated uncertainties is given in Table \ref{tab:KES_Eigenvectors} ($\rm KErSe_2$)  and Table \ref{tab:CES_Eigenvectors} ($\rm CsErSe_2$).

\begin{table*}
\caption{Eigenvectors and Eigenvalues for $\rm KErSe_2$}
\begin{ruledtabular}
\begin{tabular}{c|rrrrrrr}
Eigenvalues (meV) & &  Eigenvectors & & & & & \tabularnewline
 \hline 
0.0  & & & $ 0.52(2)~ | - \frac{13}{2}\rangle $ & $-0.508(5)~ | - \frac{7}{2}\rangle $ & $+0.58(3)~ | - \frac{1}{2}\rangle $ & $+0.347(6)~ | \frac{5}{2}\rangle $ & $+0.118(6)~ | \frac{11}{2}\rangle $  \tabularnewline
0.0  & & & $ -0.118(6)~ | - \frac{11}{2}\rangle $ & $+0.347(6)~ | - \frac{5}{2}\rangle $ & $-0.58(3)~ | \frac{1}{2}\rangle $ & $-0.508(5)~ | \frac{7}{2}\rangle $ & $-0.52(2)~ | \frac{13}{2}\rangle $  \tabularnewline
0.903(1)  & & $ 0.718(9)~ | - \frac{15}{2}\rangle $ & $-0.484(3)~ | - \frac{9}{2}\rangle $ & $+0.41(2)~ | - \frac{3}{2}\rangle $ & $+0.28(6)~ | \frac{3}{2}\rangle $ & $+0.07(7)~ | \frac{9}{2}\rangle $ & $+0.1(1)~ | \frac{15}{2}\rangle $  \tabularnewline
0.903(1)  & & $ -0.1(1)~ | - \frac{15}{2}\rangle $ & $+0.07(7)~ | - \frac{9}{2}\rangle $ & $-0.28(6)~ | - \frac{3}{2}\rangle $ & $+0.41(2)~ | \frac{3}{2}\rangle $ & $+0.484(3)~ | \frac{9}{2}\rangle $ & $+0.718(9)~ | \frac{15}{2}\rangle $  \tabularnewline
3.491(3)  & & & $ -0.7(4)~ | - \frac{13}{2}\rangle $ & $+0.1(2)~ | - \frac{7}{2}\rangle $ & $+0.7(4)~ | - \frac{1}{2}\rangle $ & $+0.1(2)~ | \frac{5}{2}\rangle $ & $-0.1(2)~ | \frac{11}{2}\rangle $  \tabularnewline
3.491(3)  & & & $ -0.083(4)~ | - \frac{11}{2}\rangle $ & $-0.1(2)~ | - \frac{5}{2}\rangle $ & $+0.68(1)~ | \frac{1}{2}\rangle $ & $-0.11(3)~ | \frac{7}{2}\rangle $ & $-0.71(9)~ | \frac{13}{2}\rangle $  \tabularnewline
5.134(3)  & & & $ -0.708(7)~ | - \frac{11}{2}\rangle $ & $+0.62(1)~ | - \frac{5}{2}\rangle $ & $+0.244(9)~ | \frac{1}{2}\rangle $ & $+0.0828(9)~ | \frac{7}{2}\rangle $ & $+0.2202(1)~ | \frac{13}{2}\rangle $  \tabularnewline
5.134(3)  & & & $ -0.2202(1)~ | - \frac{13}{2}\rangle $ & $+0.0828(9)~ | - \frac{7}{2}\rangle $ & $-0.244(9)~ | - \frac{1}{2}\rangle $ & $+0.62(1)~ | \frac{5}{2}\rangle $ & $+0.708(7)~ | \frac{11}{2}\rangle $  \tabularnewline
5.538(1)  & & $ -0.0(6)~ | - \frac{15}{2}\rangle $ & $+0.05(3)~ | - \frac{9}{2}\rangle $ & $-0.34(7)~ | - \frac{3}{2}\rangle $ & $+0.66(4)~ | \frac{3}{2}\rangle $ & $+0.2(1)~ | \frac{9}{2}\rangle $ & $-0.64(1)~ | \frac{15}{2}\rangle $  \tabularnewline
5.538(1)  & & $ -0.64(1)~ | - \frac{15}{2}\rangle $ & $-0.2(1)~ | - \frac{9}{2}\rangle $ & $+0.66(4)~ | - \frac{3}{2}\rangle $ & $+0.34(7)~ | \frac{3}{2}\rangle $ & $+0.05(3)~ | \frac{9}{2}\rangle $ & $+0.0(6)~ | \frac{15}{2}\rangle $  \tabularnewline
23.3(3)  & & & $ 0.607(1)~ | - \frac{11}{2}\rangle $ & $+0.529(5)~ | - \frac{5}{2}\rangle $ & $+0.34(2)~ | \frac{1}{2}\rangle $ & $-0.421(7)~ | \frac{7}{2}\rangle $ & $+0.2471(7)~ | \frac{13}{2}\rangle $  \tabularnewline
23.3(3)  & & & $ 0.2471(7)~ | - \frac{13}{2}\rangle $ & $+0.421(7)~ | - \frac{7}{2}\rangle $ & $+0.34(2)~ | - \frac{1}{2}\rangle $ & $-0.529(5)~ | \frac{5}{2}\rangle $ & $+0.607(1)~ | \frac{11}{2}\rangle $  \tabularnewline
25.2(3)  & & $ 0.267(9)~ | - \frac{15}{2}\rangle $ & $+0.848(4)~ | - \frac{9}{2}\rangle $ & $+0.43(1)~ | - \frac{3}{2}\rangle $ & $+0.156(8)~ | \frac{3}{2}\rangle $ & $+0.021(3)~ | \frac{9}{2}\rangle $ & $+0.0(7)~ | \frac{15}{2}\rangle $  \tabularnewline
25.2(3)  & & $ 0.0(8)~ | - \frac{15}{2}\rangle $ & $-0.021(3)~ | - \frac{9}{2}\rangle $ & $+0.156(8)~ | - \frac{3}{2}\rangle $ & $-0.43(1)~ | \frac{3}{2}\rangle $ & $+0.848(4)~ | \frac{9}{2}\rangle $ & $-0.267(9)~ | \frac{15}{2}\rangle $  \tabularnewline
25.4(3)  & & & $ -0.3(3)~ | - \frac{11}{2}\rangle $ & $-0.5(5)~ | - \frac{5}{2}\rangle $ & $+0.1(1)~ | \frac{1}{2}\rangle $ & $-0.7(7)~ | \frac{7}{2}\rangle $ & $+0.3(3)~ | \frac{13}{2}\rangle $  \tabularnewline
25.4(3)  & & & $ -0.3(3)~ | - \frac{13}{2}\rangle $ & $-0.7(7)~ | - \frac{7}{2}\rangle $ & $-0.1(1)~ | - \frac{1}{2}\rangle $ & $-0.5(5)~ | \frac{5}{2}\rangle $ & $+0.3(3)~ | \frac{11}{2}\rangle $  \tabularnewline
\end{tabular}\end{ruledtabular}
\label{tab:KES_Eigenvectors}
\end{table*}

\begin{table*}
\caption{Eigenvectors and Eigenvalues for $\rm CsErSe_2$}
\begin{ruledtabular}
\begin{tabular}{c|rrrrrr}
Eigenvalues (meV) & &  Eigenvectors & & & & \tabularnewline
 \hline 
0.0  & & $ 0.123(9)~ | - \frac{11}{2}\rangle $ & $-0.338(2)~ | - \frac{5}{2}\rangle $ & $+0.51(5)~ | \frac{1}{2}\rangle $ & $+0.513(3)~ | \frac{7}{2}\rangle $ & $+0.59(4)~ | \frac{13}{2}\rangle $  \tabularnewline
0.0  & & $ -0.59(4)~ | - \frac{13}{2}\rangle $ & $+0.513(3)~ | - \frac{7}{2}\rangle $ & $-0.51(5)~ | - \frac{1}{2}\rangle $ & $-0.338(2)~ | \frac{5}{2}\rangle $ & $-0.123(9)~ | \frac{11}{2}\rangle $  \tabularnewline
0.675(5)  & & $ 0.03(7)~ | - \frac{9}{2}\rangle $ & $-0.2(4)~ | - \frac{3}{2}\rangle $ & $+0.4(7)~ | \frac{3}{2}\rangle $ & $+0.5(9)~ | \frac{9}{2}\rangle $ & $+0.8(6)~ | \frac{15}{2}\rangle $  \tabularnewline
0.675(5)  & & $ -0.8(6)~ | - \frac{15}{2}\rangle $ & $+0.5(9)~ | - \frac{9}{2}\rangle $ & $-0.4(7)~ | - \frac{3}{2}\rangle $ & $-0.2(4)~ | \frac{3}{2}\rangle $ & $-0.03(7)~ | \frac{9}{2}\rangle $  \tabularnewline
3.29(7)  & & $ 0.68(3)~ | - \frac{13}{2}\rangle $ & $-0.04(5)~ | - \frac{7}{2}\rangle $ & $-0.71(2)~ | - \frac{1}{2}\rangle $ & $-0.17(4)~ | \frac{5}{2}\rangle $ & $+0.04(1)~ | \frac{11}{2}\rangle $  \tabularnewline
3.29(7)  & & $ 0.04(1)~ | - \frac{11}{2}\rangle $ & $+0.17(4)~ | - \frac{5}{2}\rangle $ & $-0.71(2)~ | \frac{1}{2}\rangle $ & $+0.04(5)~ | \frac{7}{2}\rangle $ & $+0.68(3)~ | \frac{13}{2}\rangle $  \tabularnewline
5.02(1)  & & $ 0.1852(6)~ | - \frac{13}{2}\rangle $ & $-0.074(3)~ | - \frac{7}{2}\rangle $ & $+0.28(2)~ | - \frac{1}{2}\rangle $ & $-0.59(2)~ | \frac{5}{2}\rangle $ & $-0.73(1)~ | \frac{11}{2}\rangle $  \tabularnewline
5.02(1)  & & $ -0.73(1)~ | - \frac{11}{2}\rangle $ & $+0.59(2)~ | - \frac{5}{2}\rangle $ & $+0.28(2)~ | \frac{1}{2}\rangle $ & $+0.074(3)~ | \frac{7}{2}\rangle $ & $+0.1852(6)~ | \frac{13}{2}\rangle $  \tabularnewline
5.28(1)  & & $ 0.06(5)~ | - \frac{9}{2}\rangle $ & $-0.4(1)~ | - \frac{3}{2}\rangle $ & $+0.67(7)~ | \frac{3}{2}\rangle $ & $+0.26(3)~ | \frac{9}{2}\rangle $ & $-0.58(2)~ | \frac{15}{2}\rangle $  \tabularnewline
5.28(1)  & & $ 0.58(2)~ | - \frac{15}{2}\rangle $ & $+0.26(3)~ | - \frac{9}{2}\rangle $ & $-0.67(7)~ | - \frac{3}{2}\rangle $ & $-0.4(1)~ | \frac{3}{2}\rangle $ & $-0.06(5)~ | \frac{9}{2}\rangle $  \tabularnewline
23.1(5)  & & $ -0.239(2)~ | - \frac{13}{2}\rangle $ & $-0.428(8)~ | - \frac{7}{2}\rangle $ & $-0.36(3)~ | - \frac{1}{2}\rangle $ & $+0.531(7)~ | \frac{5}{2}\rangle $ & $-0.59(2)~ | \frac{11}{2}\rangle $  \tabularnewline
23.1(5)  & & $ 0.59(2)~ | - \frac{11}{2}\rangle $ & $+0.531(7)~ | - \frac{5}{2}\rangle $ & $+0.36(3)~ | \frac{1}{2}\rangle $ & $-0.428(8)~ | \frac{7}{2}\rangle $ & $+0.239(2)~ | \frac{13}{2}\rangle $  \tabularnewline
24.9(5)  & & $ -0.25(2)~ | - \frac{15}{2}\rangle $ & $-0.843(8)~ | - \frac{9}{2}\rangle $ & $-0.45(2)~ | - \frac{3}{2}\rangle $ & $-0.17(1)~ | \frac{3}{2}\rangle $ & $-0.025(6)~ | \frac{9}{2}\rangle $  \tabularnewline
24.9(5)  & & $ 0.025(6)~ | - \frac{9}{2}\rangle $ & $-0.17(1)~ | - \frac{3}{2}\rangle $ & $+0.45(2)~ | \frac{3}{2}\rangle $ & $-0.843(8)~ | \frac{9}{2}\rangle $ & $+0.25(2)~ | \frac{15}{2}\rangle $  \tabularnewline
25.2(6)  & & $ -0.32(2)~ | - \frac{13}{2}\rangle $ & $-0.74(6)~ | - \frac{7}{2}\rangle $ & $-0.136(1)~ | - \frac{1}{2}\rangle $ & $-0.47(2)~ | \frac{5}{2}\rangle $ & $+0.3262(5)~ | \frac{11}{2}\rangle $  \tabularnewline
25.2(6)  & & $ -0.3262(5)~ | - \frac{11}{2}\rangle $ & $-0.47(2)~ | - \frac{5}{2}\rangle $ & $+0.136(1)~ | \frac{1}{2}\rangle $ & $-0.74(6)~ | \frac{7}{2}\rangle $ & $+0.32(2)~ | \frac{13}{2}\rangle $  \tabularnewline
\end{tabular}\end{ruledtabular}
\label{tab:CES_Eigenvectors}
\end{table*}

%


\end{document}